# A Computation Control Motion Estimation Method for Complexity-Scalable Video Coding

Weiyao Lin, Krit Panusopone, David M. Baylon, and Ming-Ting Sun

***Abstract*—In this paper, a new Computation-Control Motion Estimation (CCME) method is proposed which can perform Motion Estimation (ME) adaptively under different computation or power budgets while keeping high coding performance. We first propose a new class-based method to measure the Macroblock (MB) importance where MBs are classified into different classes and their importance is measured by combining their class information as well as their initial matching cost information. Based on the new MB importance measure, a complete CCME framework is then proposed to allocate computation for ME. The proposed method performs ME in a one-pass flow. Experimental results demonstrate that the proposed method can allocate computation more accurately than previous methods and thus has better performance under the same computation budget.**

## I. Introduction And Related Work

COMPLEXITY-Scalable Video Coding (CSVC) (or Computational-Scalable/Power-aware video coding) is of increasing importance to many applications [1-5,11,13,14,18], such as video communication over mobile devices with limited power budget as well as real-time video systems which require coding the video below a fixed number of processor computation cycles.

The target of the CSVC research is to find an efficient way to allocate the available computation budget for different Video Parts (e.g., Group of Pictures (GOPs), frames, and Macroblocks (MBs)) and different Coding Modules (e.g., Motion Estimation (ME), Discrete Cosine Transform (DCT), and Entropy Coding) so that the resulting video quality is kept as high as possible under the given computation budget. Since the available computation budget may vary, the CSVC algorithm should be able to perform video coding under different budget levels.

Since ME occupies the major portion of the whole coding complexity [6,12], we will focus on the computation allocation for the ME part in this paper (i.e., Computation-Control Motion Estimation (CCME)). Furthermore, since the computation often can be roughly measured by the number of Search Points (SPs) in ME, we will use the term *SP* and *Computation* interchangeably.

Many algorithms have been proposed for CCME [1-5,14]. They can be evaluated by two key parts of CCME: (1) *the computation allocation,* and (2) *the MB importance measure*. They are described as follows.

(1) *The computation allocation order.* Two approaches can be used for allocating the computations: one-pass flow and multi-pass flow. Most previous CCME methods [2-4] allocate computation in a multi-pass flow, where MBs in one frame are processed in a step-by-step fashion based on a table which measures the MB importance. At each step, the computation is allocated to the MB that is measured as *the most important* among all the MBs in the whole frame. The table is updated after each step. Since the multi-pass methods use a table for all MBs in the frame, they can have a global view of the whole frame while allocating computation. However, they do not follow the regular coding order and require the ME process to jump between MBs, which is less desirable for hardware implementations. Furthermore, since the multi-pass methods do not follow the regular coding order, the neighboring MB information cannot be used for prediction to achieve better performance. Compared to the multi-pass flow approach, one-pass methods [5,14] allocate computation and perform ME in the regular video coding order. They are more favorable for hardware implementation and can also utilize the information from neighboring MBs. However, it is more difficult to develop a good one-pass method since (a) a one-pass method lacks a global view of the entire frame and may allocate unbalanced computations to different areas of the frame, and (b) it is more difficult to find a suitable method to measure the importance of MBs.

(2) *The MB importance measure*. In order to allocate computation efficiently to different MBs, it is important to measure the importance of the MBs for the coding performance, so that more computation will be allocated to the more *important* MBs (i.e., MBs with larger importance measure values). Tai et al. [2] use the current *Sum of Absolute Difference* (*SAD)* value for the MB importance measure. Their assumption is that MBs with large matching costs will have more room to improve, and thus more search points will be allocated to these MBs. Chen et al. [5,14] use a similar measure in their one-pass method. However, the assumption that larger current SAD will lead to bigger SAD decrease is not always guaranteed, which makes the allocation less accurate. Yang et al. [3] use the ratio between the *SAD decrease* and *the number of SPs* at the previous ME step to measure the MB importance.

Weiyao Lin is with the Institute of Image Communication and Information Processing, Shanghai Jiao Tong University, Shanghai, China (e-mail: hellomikelin@gmail.com).

Krit Panusopone and David M. Baylon are with the Advanced Technology Department, CTO Office, Home & Networks Mobility, Motorola Inc.

Ming-Ting Sun is with the Department of Electrical Engineering, University of Washington, Seattle, USA (e-mail: mts@u.washington.edu).

Kim et al. [4] use a similar measure except that they use *Rate-Distortion Cost Decrease* [4] instead of the *SAD decrease*. However, their methods can only be used in multi-pass methods where the allocation is performed in a step-by-step fashion and cannot be applied to one-pass methods.

In this paper, a new one-pass CCME method is proposed. We first propose a Class-based MB Importance Measure (CIM) method where MBs are classified into different classes based on their properties. The importance of each MB is measured by combining its class information as well as its initial matching cost value. Based on the CIM method, a complete CCME framework is then proposed which first divides the total computation budget into independent sub-budgets for different MB classes and then allocates the computation from the class budget to each step of the ME process. Furthermore, the proposed method performs ME in a one-pass flow, which is more desirable for hardware implementation. Experimental results demonstrate that the proposed method can allocate computation more accurately than previous methods while maintaining good quality.

The rest of the paper is organized as follows: Section II describes our proposed CIM method. Based on the CIM method, Section III describes the proposed CCME algorithm in detail. The experimental results are given in Section IV. Section V gives some discussions, and Section VI concludes the paper.

## II. THE CLASS-BASED MB IMPORTANCE MEASURE

In this section, we discuss some statistics of ME and describe our Class-based MB Importance Measure method in detail. For convenience, we use COST [10] as the ME matching cost in the rest of the paper. The COST [10] is defined as in Eqn. (1):

$$COST = SAD + \lambda_{MOTION} \cdot R(MV) \tag{1}$$

where *SAD* is the Sum of Absolute Difference for the block matching error, $R(MV)$ is the number of bits to code the *Motion Vector* (*MV*), and $\lambda_{MOTION}$ is the Lagrange multiplier [19].

In this paper, the CIM method and the proposed CCME algorithm is described based on the Simplified Hexagon Search (SHS) [7] algorithm. However, our algorithms are general and can easily be extended to other ME algorithms [9,10,15-17].

The SHS is a newly developed ME algorithm which can achieve performance close to Full Search (FS) with comparatively low SPs. The SHS process can be described as in Fig. 1.

Before the ME process, the SHS algorithm first checks the *init_COST* which is defined as:

$$init_COST = \min\left(COST_{(0,0)}, COST_{PMV}\right) \tag{2}$$

where $COST_{(0,0)}$ is the COST of the (*0*,*0*) *MV*, and $COST_{PMV}$ is the COST of the *Predictive MV* (*PMV*) [7]. If *init_COST* is smaller than a threshold $th_1$, the SHS algorithm will stop after performing a *small local search* (search 4 points around the position of the *init_COST*), which we call *the Upper Path*. If *init_COST* is larger than the threshold, the SHS algorithm will proceed to the steps of *Small Local Search*, *Cross Search*, *Multiple Hexagon Search*, *Small Hexagon Search* and *Small Diamond Search* [7], which we call *the Lower Path*. Inside the lower path, another threshold $th_2$ is used to decide whether or not to skip the steps of *Cross Search* and *Multi Hexagon Search*.

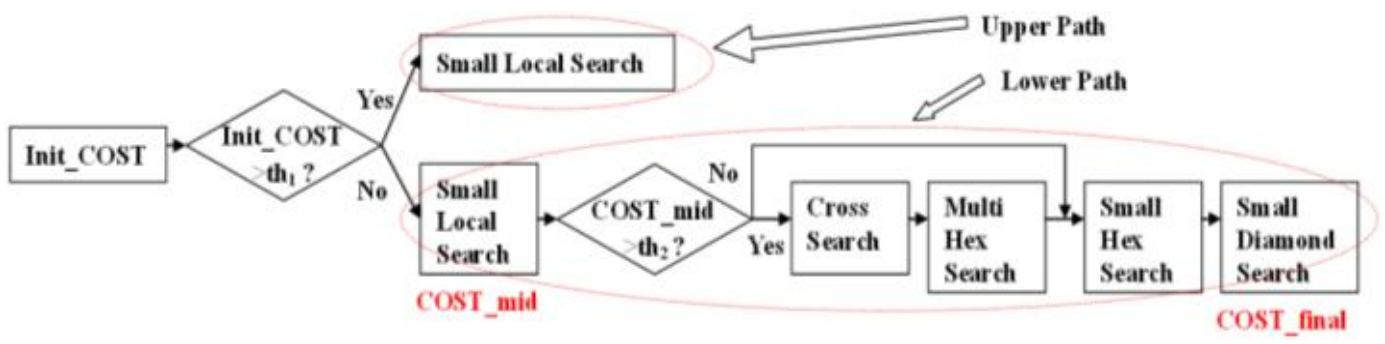

Fig 1. The SHS process.

### A. Analysis of Motion Estimation Statistics

In order to analyze the relationship between the COST value and the number of search points, we define two more COSTs: *COST_mid* (the COST value right after the *Small Local Search* step in the *Lower Path*) and *COST_final* (the COST value after going through the entire ME process), as in Fig. 1. Three MB classes are defined as:

$$Class_{cur_MB} = \begin{cases} 1 & \text{if } init_COST < th_1 \\ 2 & \text{if } init_COST \geq th_1 \text{ and } |COST_mid - COST_final| > c \\ 3 & \text{if } init_COST \geq th_1 \text{ and } |COST_mid - COST_final| \leq c \end{cases} \tag{3}$$

where *cur_MB* is the current MB, $th_1$ is the threshold defined in the SHS algorithm [7] to decide whether the *init_COST* is large or small [7], and *c* is another threshold to decide the significance of the cost improvement between *COST_mid* and *COST_final*. MBs in Class 1 are MBs with small current COST values. Class 2 represents MBs with large current COST values where additional searches can yield significant improvement. Class 3 represents MBs with large current COST values but where further searches do not produce significant improvement. If we can predict Class 3 MBs, we can save computation by skipping further searches for the Class 3 MBs. It should be noted that since we cannot get *COST_final* before actually going through the *Lower Path*, the classification method of Eqn. (3) is only used for statistical analysis. A practical classification method will be proposed later in this section. Furthermore, since MBs in Class 1 have small current COST value, their MB importance measure can be easily defined. Therefore, we will focus on the analysis of Class 2 and Class 3 MBs.

Table 1 lists the percentage of Class 1, Class 2 and Class 3 MBs over the total MBs for sequences of different resolutions and under different Quantization Parameter (QP) values where *c* of Eqn. (3) is set to be different values of *0*, *2% of COST_mid*, and *4% of COST_mid*. It should be noted that *0* is the smallest possible value for *c*. We can see from Table 1 that the number of Class 3 MBs will become even larger if *c* is relaxed to larger values.

Table 1
Percentage of Class 1, Class 2, and Class 3 MBs over the total MBs (100 frames for Qcif and 50 frames for Cif and SD)

| Sequence | | | QP=23 | | | QP=28 | | | QP=33 | | |
|---|---|---|---|---|---|---|---|---|---|---|---|
| | | | Class 1 MB | Class 2 MB | Class 3 MB | Class 1 MB | Class 2 MB | Class 3 MB | Class 1 MB | Class 2 MB | Class 3 MB |
| Qcif (176x144) | Foreman_Qcif (*c=0*) | | 50% | 5.5% | 44.4% | 33.8% | 6.7% | 59.4% | 14.9% | 8.2% | 76.7% |
| | Akiyo_Qcif (*c=0*) | | 96% | 0% | 4% | 89% | 0% | 10% | 68.7% | 0% | 31.2% |
| | Mobile_Qcif (*c=0*) | | 6.9% | 0.7% | 92.2% | 1.5% | 0.8% | 97.6% | 0.6% | 0.8% | 98.4% |
| Cif (352x288) | Bus_Cif | *c=0* | 21.6% | 21.8% | 56.8% | 14.6% | 22.2% | 63.1% | 4.2% | 25.7% | 70% |
| | | *c=2% Cost_mid* | 21.6% | 20.5% | 57.9% | 14.6% | 20.8% | 64.6% | 4.2% | 22.9% | 72.8% |
| | | *c=4% Cost_mid* | 21.6% | 19.5% | 58.9% | 14.6% | 19.4% | 66% | 4.2% | 20.6% | 75.1% |
| | Football_Cif (*c=0*) | | 22.4% | 53.1% | 24.5% | 15.3% | 54.1% | 30.5% | 2.3$ | 58% | 39.7% |
| | Container_Cif (*c=0*) | | 90.6% | 0% | 9.3% | 65.6% | 0.2% | 34.2% | 48.8% | 2.6% | 48.6% |
| | Mobile_Cif | *c=0* | 11% | 8.1% | 80.9% | 7.2% | 8.5% | 84.3% | 4.3% | 9.7% | 86% |
| | | *c=2% Cost_mid* | 11% | 7.3% | 81.7% | 7.2% | 7.7% | 85.1% | 4.3% | 8.4% | 87.3% |
| | | *c=4% Cost_mid* | 11% | 6.6% | 82.4% | 7.2% | 6.8% | 86% | 4.3% | 7.3% | 88.4% |
| | Foreman_Cif (*c=0*) | | 61.6% | 12% | 26.4% | 51.5% | 13.3% | 35.2% | 32.9% | 17.1% | 50% |
| SD (720x576) | Mobile_SD (*c=0*) | | 37.6% | 7.4% | 55% | 22.5% | 7.9% | 69.6% | 12% | 9% | 79% |
| | Football_SD (*c=0*) | | 41.7% | 29.4% | 28.9% | 32% | 30% | 38% | 20.1% | 32.1% | 47.8% |
| | Flower_SD (*c=0*) | | 28.7% | 8.7% | 62.6% | 25.1% | 9.6% | 65.3% | 22.7% | 11.4% | 65.9% |

Fig. 2 shows the COST value distribution of Class 2 MBs and Class 3 MBs where *c* of Eqn. (3) is set to be *0*. We only show results for Foreman_qcif with QP=28 in Fig. 2. Similar results can be observed for other sequences and other QP values. In Fig. 2, 20 frames are coded. The experimental setting is the same as that described in Section 5. In order to have a complete observation, all the three COST values are displayed in Fig. 2, where Fig. 2(a), Fig. 2(b) and Fig. 2(c) show the distributions of *init_COST*, *COST_mid*, and *COST_final* respectively.

From Fig. 2 and Table 1, we can observe that (a) a large portion of MBs with large current COST values can be classified as Class 3 where only a few SPs are needed and additional SPs do not produce significant improvement, and (b) The distribution of all the three COSTs for Class 2 and Class 3 are quite similar. This implies that Class 2 or Class 3 cannot be differentiated based on their COST value only.

Based on the above observations, we can draw several conclusions for the computation allocation as follows:

(1) The number of SPs needed for keeping the performance for each MB is not always related to its current COST value. Therefore, using the COST value only as the MB importance measure, which is used by many previous methods [3,5,14], may not allocate SPs efficiently.

(2) Further experiments show that for Class 2 MBs, the number of SPs needed for keeping the performance is roughly proportional to their *init_COST* value (although it is not true if Class 2 and Class 3 MBs are put together).

These imply that we can have a better MB importance measure if we use the class and COST information together.

As mentioned, since we cannot get *COST_final* before going through the Lower Path, Class 2 and Class 3 cannot be differentiated by their definition in Eqn. (3) in practice. Furthermore, since the COST distribution of Class 2 and Class 3 is similar, the current COST value cannot differentiate between these two classes. Therefore, before describing our MB Importance Measure method, we first propose a practical MB classification method which we call the Predictive-MV-Accuracy-based Classification (PAC) algorithm. The PAC algorithm will be described in the following section.

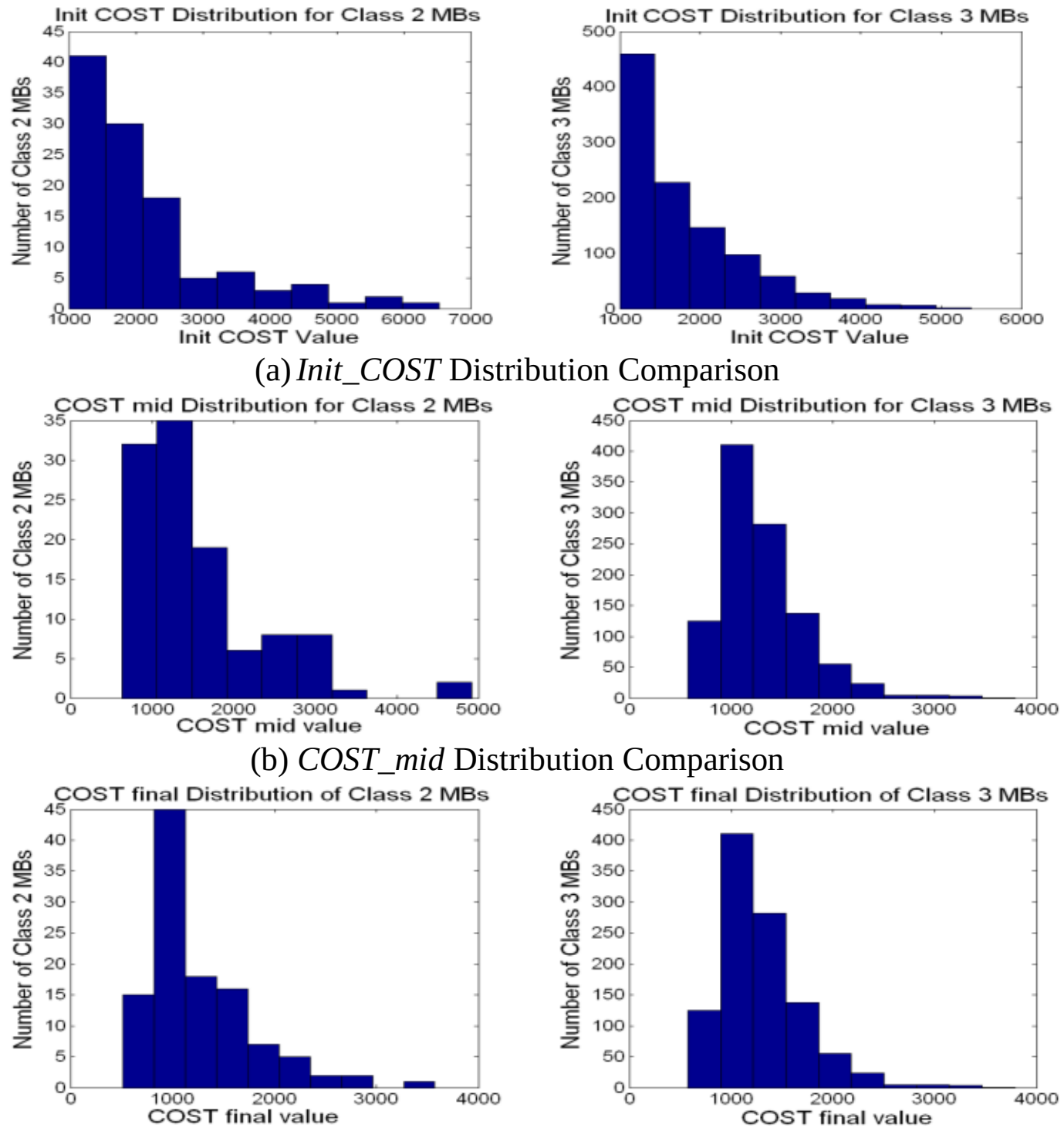

(c) *COST_final* Distribution Comparison

Fig 2. COST value distribution for class 2 and class 3 MBs for Foreman_qcif sequence (Left: Class 2, right: Class 3).

## *B. The Predictive-MV-Accuracy-based Classification Algorithm*

The proposed PAC algorithm converts the definition of Class 2 and Class 3 from the *COST value* point of view to the *Predictive MV accuracy* point of view.

The basic idea of the PAC algorithm is described as follows: (1) If the motion pattern of a MB can be predicted accurately

(i.e., if PMV is accurate), then only a small local search is needed to find the *final MV* (i.e., the MV of *COST_final*). In this case, no matter how large the COST is, additional search points after the small local search are not needed because the *final MV* has already been found by the small local search. This corresponds to Class 3 MBs.

(2) On the other hand, if the motion pattern of a MB cannot be accurately predicted, a small local search will not be able to find the *final MV*. In this case, a large area search (i.e., the *Lower Path*) after the small local search is needed to find the *final MV* with a lower COST value. This corresponds to Class 2 MBs.

Since the *MV_final* (MV for *COST_final*) cannot be obtained before going through the *Lower Path*, the final MV of the co-located MB in the previous frame is used instead to measure the accuracy of motion-pattern prediction. Therefore, the proposed PAC algorithm can be described as:

$$Class_{cur_MB} = \begin{cases} 1 & if\ init_COST < th \\ 2 & if\ init_COST \geq th\ and\ |PMV_{cur_MB} - MV_{pre_final}| > Th \\ 3 & if\ init_COST \geq th\ and\ |PMV_{cur_MB} - MV_{pre_final}| \leq Th \end{cases} \quad (4)$$

where $|PMV_{cur_MB}$-$MV_{pre_final}|$ is the measure of the motion-pattern-prediction accuracy, $PMV_{cur_MB}$ is the PMV [7] of the current MB, $MV_{pre_final}$ is the final MV of the co-located MB in the previous frame, and *Th* is the threshold to check whether the *PMV* is accurate or not. *Th* can be defined based on different small local search patterns. In the case of SHS, *Th* can be set as 1 in integer pixel resolution. According to Eqn. (4), Class 1 includes MBs that can find good matches from the previous frames. MBs with irregular or unpredictable motion patterns will be classified as Class 2. Class 3 MBs will include areas with complex textures but similar motion patterns to the previous frames.

It should be noted that the classification using Eqn. (4) is very tight (in our case, any MV difference larger than 1 integer pixel will be classified as Class 2 and a large area search will be performed). Furthermore, by including $MV_{pre_final}$ for classification, we also take the advantage of including the temporal motion-smoothness information when measuring motion-pattern-prediction accuracy. Therefore, it is reasonable to use $MV_{pre_final}$ to take the place of *MV_final*. This will be demonstrated in Table 2 and Fig. 3 in the following and will be further demonstrated in the experimental results.

Table 2 The detection rates of the PAC algorithm.

| Sequence | Class 2 Detection Rate | Class 3 Detection Rate |
|---|---|---|
| Mobile Qcif | 80% | 82% |
| Football_Cif | 71% | 90% |
| Foreman_Qcif | 75% | 76% |

Table 2 shows the detection rates for Class 2 and Class 3 MBs with our PAC algorithm for some sequences, where the class definition in Eqn. (3) is used as the ground truth and *c* in Eqn. (3) is set to be 0. Table 2 shows that our PAC algorithm has high MB classification accuracy.

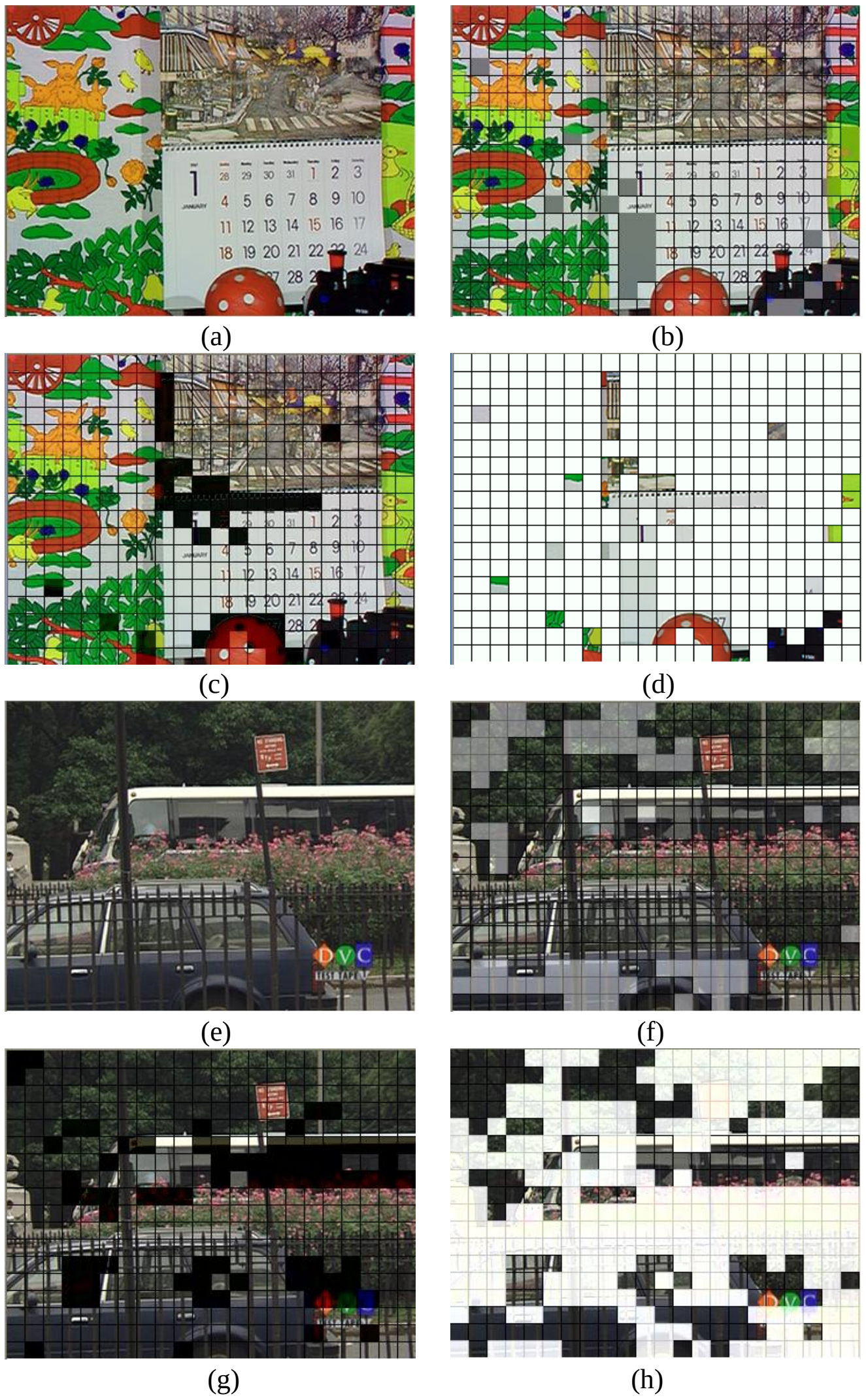

Fig. 3. The original frames (a, e) and the distributions of Class 1 (b, f), Class 2 (c, g), and Class 3 (d, h) MBs for *Mobile_Cif* and *Bus_Cif*.

Fig. 3 shows the distribution of MBs for each class of two example frames by using our PAC algorithm. Fig. 3 (a) and (e) are the original frames. Blocks labeled grey in (b) and (f) are MBs belonging to Class 1. Blocks labeled black in (c) and (g) and blocks labeled white in (d) and (h) are MBs belonging to Class 2 and Class 3, respectively.

Fig. 3 shows the reasonableness of the proposed PAC algorithm. From Fig. 3, we can see that most Class 1 MBs include backgrounds or flat areas that can find good matches in the previous frames ((b) and (f)). Areas with irregular or unpredictable motion patterns are classified as Class 2 (for example, the *edge between the calendar and the background* as well as *the bottom circling ball* in (c), and *the running bus* as well as *the down-right logo* in (g)). Most complex-texture areas are classified as Class 3, such as the complex background and calendar in (d) as well as the flower area in (h).

### C. The MB Importance Measure

Based on the discussion above and the definition of MB classes in Eqn. (4), we can describe our proposed CIM method as follows:

(1) MBs in Class 1 will always be allocated a fixed small number of SPs.
(2) MBs in Class 2 will have high importance. They will be allocated more SPs, and each Class 2 MB will have a guaranteed minimum SPs for coding performance purposes. If two MBs both belong to Class 2, their comparative importance is proportional to their *init_COST* value and the SPs will be allocated accordingly.
(3) MBs in Class 3 will have lower importance than MBs in Class 2. Similar to Class 2, we make the comparative importance of MBs within Class 3 also proportional to their *init_COST* value. By allowing some Class 3 MBs to have more SPs rather than fixing the SPs for each MB, the possible performance decrease due to the mis-classification of MBs from Eqn. (4) can be avoided. This will be demonstrated in the experimental results.

With the CIM method, we can have a more accurate MB importance measure by differentiating MBs into classes and combining the class and the COST information. Based on the CIM method, we can develop a more efficient CCME algorithm. The proposed CCME algorithm will be described in detail in the following section.

## III. The CCME Algorithm

The framework of the proposed CCME algorithm is described in Fig. 4.

From Fig. 4, the proposed CCME algorithm has four steps:
(1) *Frame Level computation Allocation* (*FLA*). Given the available total computation budget for the whole video sequence, FLA allocates a computation budget to each frame.
(2) *Class Level computation Allocation* (*CLA*). After one frame is allocated a computation budget, CLA further divides the computation into three independent sub-budgets (or class budgets) with one budget for each class defined in Eqn. (4).
(3) *MB Level computation Allocation* (*MLA*). When performing ME, each MB will first be classified into one of the three classes according to Eqn. (4). MLA then allocates the computation to the MB from its corresponding class budget.
(4) *Step Level computation Allocation* (*SLA*). After an MB is allocated a computation budget, SLA allocates these computations into each ME step.

It should be noted that the CLA step and the MLA step are the key steps of the proposed CCME algorithm where our proposed CIM method is implemented. Furthermore, we also investigated two strategies for computation allocation for CLA and MLA steps: *the tight strategy* and *the loose strategy*. For the tight strategy, the actual computation used in the current frame *must be lower* than the computation allocated to this frame. Due to this property, the FLA step is sometimes not necessary for the tight strategy. In some applications, we can simply set the budget for all frames as a fixed number for performing the tight strategy. For the *loose strategy*, the actual computation used *for some frames* can exceed the computation allocated to these frames but the total computation used for the *whole sequence* must be lower than the budget. Since the loose strategy allows frames to borrow computation from others, the FLA step is needed to guarantee that the total computation used for *the whole sequence* will not exceed the available budget.

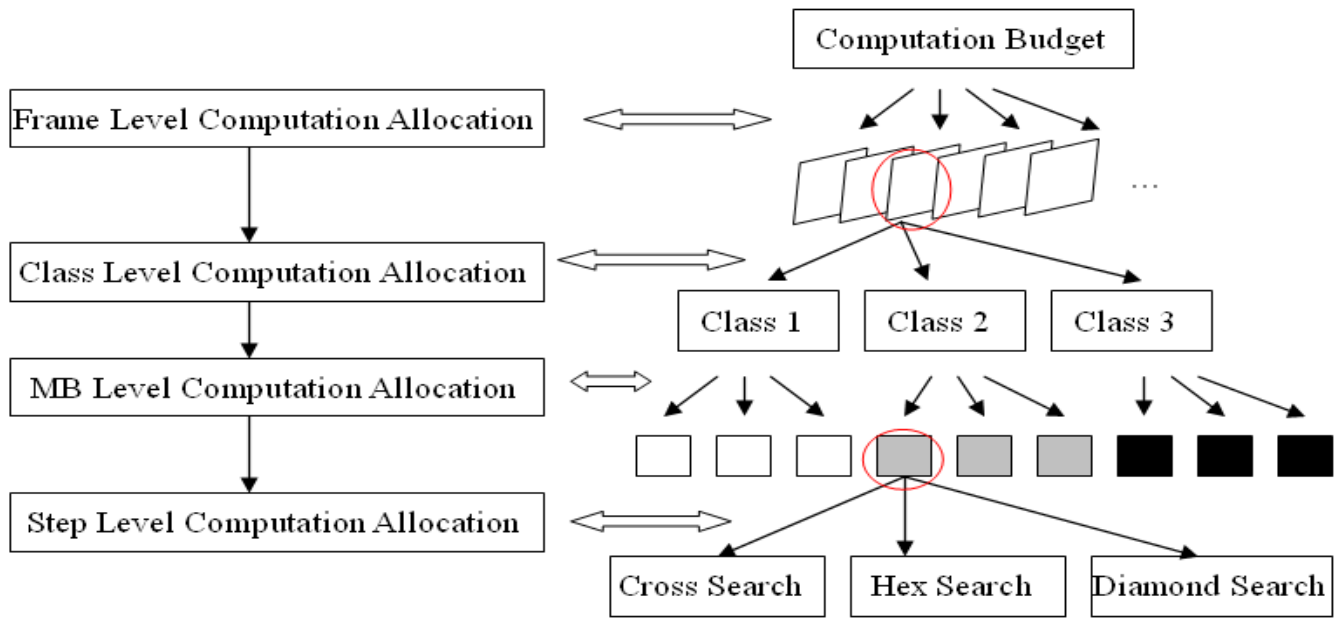

Fig. 4 The framework for the proposed CCME algorithm.

Since the performances of the loose-strategy algorithm and the tight-strategy algorithm are similar based on our experiments, we will only describe our algorithm based on the tight strategy in this paper. It should be noted that since the basic ideas of the CLA and MLA processes are similar for both the tight and loose strategies, a loose-strategy algorithm can be easily derived from the description in this paper. Furthermore, as mentioned, the FLA step is sometimes unnecessary for the tight strategy. In order to prevent the effect of frame level allocation and to have a fair comparison with other methods, we also skip the FLA step by simply fixing the target computation budget for each frame in this paper. In practice, various frame-level allocation methods [2-5] can be easily incorporated into our algorithm.

### A. Class Level computation Allocation (CLA)

The basic ideas of the CLA process can be summarized as follows:
(a) In the CLA step, the computation budget for the whole frame $C_F$ is divided into three independent class budgets (i.e. $C_{Class(1)}$, $C_{Class(2)}$ and $C_{Class(3)}$). MBs from different classes will be allocated computation from their corresponding class budget and will not affect each other.
(b) Since the CLA step is based on the tight strategy in this paper, the basic layer $BL_{Class(i)}$ is first allocated to guarantee that each MB has a minimum number of SPs. The remaining SPs are then allocated to the additional layer $AL_{Class(i)}$. The total budget for each class consists of the basic layer plus the additional layer. Furthermore, since the MBs in class 1 only performs a local search, the budget for class 1 only contains the basic layer (i.e. $C_{Class(1)}= BL_{Class(1)}$ and $AL_{Class(1)}=0$).
(c) The actual computation used for each class in the previous frame ($CA^{pre}_{class(i)}$) is used as the ratio parameter for class budget allocation for the additional layer.

Therefore, the CLA process can be described as in Eqn. (5) and Fig. 5.

$$C_{class(i)} = BL_{class(i)} + AL_{class(i)} \qquad i = 1,2,3 \tag{5}$$

where $BL_{class(i)} = BL_{MB_class(i)} \cdot NM^{pre}_{class(i)}$

$$AL_{class(i)} = \begin{cases} 0 & if\ i=1 \\ \min\left( AL_F \cdot \dfrac{CA^{pre}_{class(2)}}{CA^{pre}_{class(2)} + CA^{pre}_{class(3)}},\ AL_{MB_\max_class(2)} \cdot NM^{pre}_{class(i)} \right) & if\ i=2 \\ AL_F - AL_{class(2)} & if\ i=3 \end{cases}$$

$BL_F = (BL_{class(1)} + BL_{class(2)} + BL_{class(3)})$ , $AL_F = C_F - BL_F$ , $C_{class(i)}$ is the computation allocated to class $i$, and $BL_{class(i)}$ and $AL_{class(i)}$ represent the computation allocation for the class $i$ basic layer and additional layer, respectively. $C_F$ is the total computation budget for the whole frame, and $BL_F$ and $AL_F$ represent the basic-layer computation and the additional-layer computation for the whole frame, respectively. $NM^{pre}_{class(i)}$ is the total number of MBs belonging to Class $i$ in the previous frame and $CA^{pre}_{class(i)}$ is the number of computation actually used for the Class $i$ in the previous frame. $BL_{MB_class(i)}$ is the minimum number of computations guaranteed for each MB in the basic layer. In the case of SHS, we set $BL_{MB_class(1)} = BL_{MB_class(3)} = 6$ SPs for Class 1 and Class 3, and $BL_{MB_class(2)} = 25$ SPs for Class 2. As mentioned, since Class 2 MBs have higher importance in our CIM method, we guarantee them a higher minimum SP. Furthermore, in order to avoid too many useless SPs allocated to Class 2 MBs, a maximum number of SPs ($AL_{MB_max_class(2)}$) is set. SPs larger than $AL_{MB_max_class(2)}$ are likely wasted and therefore are allocated to Class 3 MBs ($AL_F - AL_{class(2)}$).

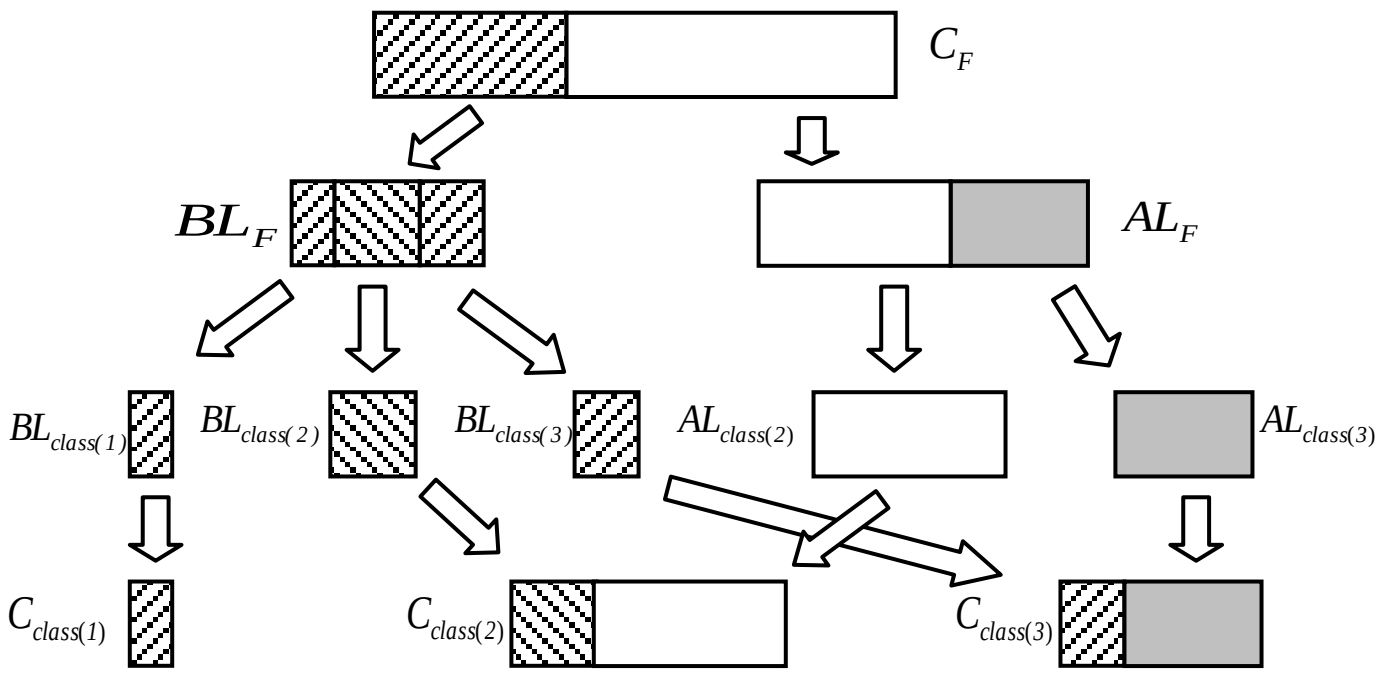

Fig. 5 The Tight-Strategy-Based CLA process.

From Eqn. (5) and Fig. 5, we can summarize several features of our CLA process as follows:

(a) Since *Class* is newly defined in this paper, the CLA step is unique in our CCME method and is not included in the previous CCME algorithms [1-5,14].

(b) When performing CLA, the information from the previous frame ($NM^{pre}_{class(i)}$ and $CA^{pre}_{class(i)}$) is used. $NM^{pre}_{class(i)}$ provides a global-view estimation of the MB class distribution for the current frame, and $CA^{pre}_{class(i)}$ is used as a ratio parameter for class budget allocation for the additional layer.

(c) The CIM method is implemented in the CLA process where (i) the *CA* for Class 2 is normally larger than other classes, and (ii) Class 2 MBs have a larger guaranteed minimum number of SPs (i.e., $BL_{MB_class(2)}$ in the Tight-SLA).

## *B. MB Level computation Allocation (MLA)*

The MLA process can be described in Eqn. (6). Similar to the CLA process, a basic layer ($BL_{MB}$) and an additional layer ($AL_{MB}$) are set. When allocating the additional layer computation, the initial COST of the current MB ($COST^{init}_{cur_MB}$) is used as a parameter to decide the number of computation allocated. The MLA process for Class 2 or Class 3 MBs is described as in Fig. 6.

$$C_{cur_MB} = BL_{cur_MB} + AL_{cur_MB} \tag{6}$$

where $BL_{cur_MB} = \begin{cases} BL_{MB_class(1)} & if\ class_{cur_MB} = 1 \\ BLC_{MB_class(2)} & if\ class_{cur_MB} = 2 \\ BL_{MB_class(3)} & if\ class_{cur_MB} = 3 \end{cases}$

$$AL_{cur_MB} = \begin{cases} 0 & if\ class_{cur_MB} = 1 \\ min\left( max\left( \dfrac{COST^{init}_{cur_MB}}{Avg_COST^{init}_{class(2)}} \cdot \dfrac{ab_{class(2)}}{nm^{pre}_{class(2)}}, 0 \right), AL_{MB_max_class(2)} \right) & if\ class_{cur_MB} = 2 \\ min\left( max\left( \dfrac{COST^{init}_{cur_MB}}{Avg_COST^{init}_{class(3)}} \cdot \dfrac{ab_{class(3)}}{nm^{pre}_{class(3)}}, 0 \right), AL_{MB_max_class(3)} \right) & if\ class_{cur_MB} = 3 \end{cases}$$

$C_{cur_MB}$ is the computation allocated to the current MB, $COST^{init}_{cur_MB}$ is the initial COST of the current MB as in Eqn. (2), $Avg_COST^{init}_{class(i)}$ is the average of the initial COST for all the already-coded MBs belonging to Class $i$ in the current frame. "$ab_{class(i)}$" is the computation budget available in the additional layer for class $i$ before coding the current MB and "$nm^{pre}_{class(i)}$" is the estimated number of remaining–uncoded MBs for class $i$ before coding the current MB. $BLC_{MB_class(2)}$ is equal to $BL_{MB_class(2)}$ if either $ab_{class(2)}>0$ or $nm_{class(2)} >1$, and equal to $BL_{MB_class(3)}$ otherwise. It should be noted that $BLC_{MB_class(2)}$ is defined to follow the tight strategy where a larger ML-BL budget ($BL_{MB_class(2)}$) is used if the available budget is sufficient and a smaller ML-BL budget ($BL_{MB_class(3)}$) is used otherwise. $AL_{MB_max_class(2)}$ and $AL_{MB_max_class(3)}$ are the same as in Eqn. (5) and are set in order to avoid too many useless SPs allocated to the current MB. In the experiments of this paper, we set $AL_{MB_max_class(i)} + BL_{MB_class(i)} = 250$ for a search range of $\pm 32$ pixels. It should be noted that since we cannot get the exact number of remaining MBs for each class before coding the whole frame, $nm^{pre}_{class(i)}$ is estimated by the parameters of the previous frame. "$ab_{class(i)}$" and "$nm^{pre}_{class(i)}$" are set as $AL_{class(i)}$ and $NM^{pre}_{class(i)}$ respectively at the beginning of each frame and are updated before coding the current MB as in Eqn. (7).

$$\begin{cases} ab_{class(i)} = ab_{class(i)} - (CA_{pre_MB} - BL_{pre_MB}) & if\ class_{pre_MB} = i \\ nm^{pre}_{class(i)} = nm^{pre}_{class(i)} - 1 & if\ class_{pre_MB} = i \end{cases} \tag{7}$$

where the definition of $AL_{class(i)}$ and $NM^{pre}_{class(i)}$ are the same as in Eqn. (5), and $CA_{pre_MB}$ and $BL_{pre_MB}$ represent the actual computation consumed and the basic layer computation allocated for the MB right before the current MB, respectively.

From Eqn. (5-7), we can see that the CLA and MLA steps are based on classification using our CIM method, where Class 1 MBs are always allocated a fixed small number of SPs, and Class 2 and Class 3 MBs are first separated into independent class budgets and then allocated based on their *init_COST* value within each class budget. Thus, the proposed CCME algorithm can combine the class information and COST information for a more precise computation allocation.

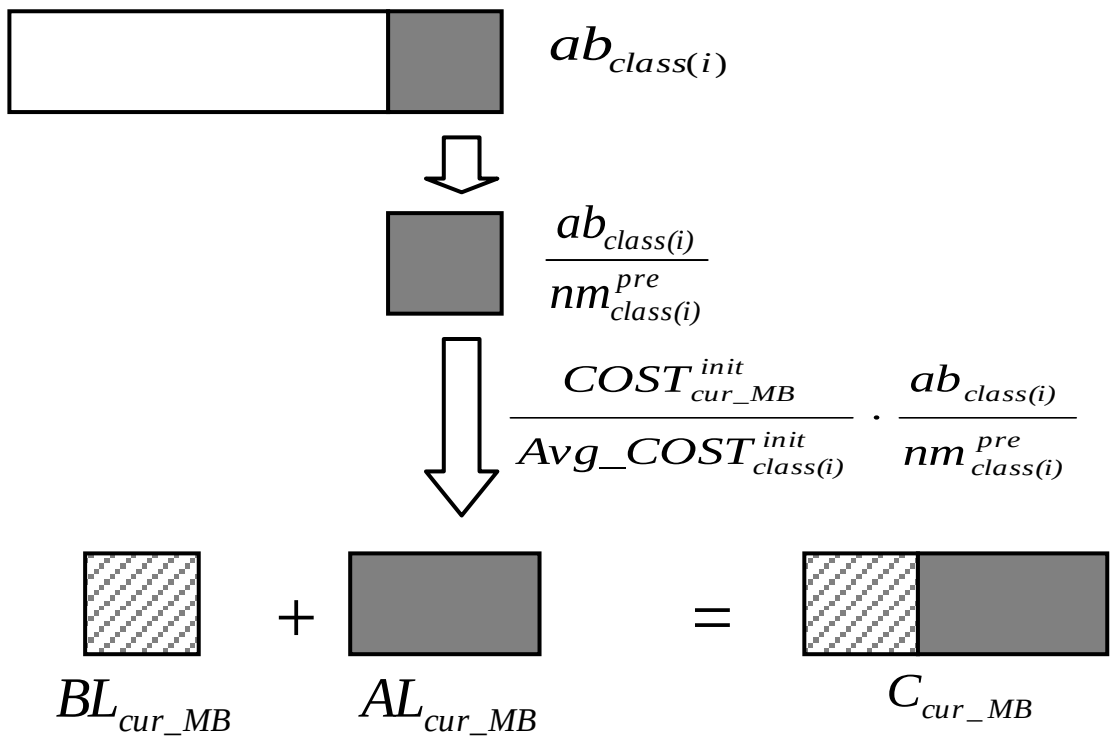

Fig. 6 The Tight-MLA process for Class 2 and Class 3 MBs.

## C. Step Level computation Allocation (SLA)

The SLA process will allocate the computation budget for an MB into each ME step. Since the SHS method is used to perform ME in this paper, we will describe our SLA step based on the SHS algorithm. However, our SLA method can easily be applied to other ME algorithms [9,10,15-17].

The SLA process can be described as in Eqn. (8).

$$\begin{cases} C_{Small_Local_Search} = C_{Step_min} \\ C_{Cross_Search} = NS_{Cross_Search} \cdot CS_{Cross_Search} \\ C_{Multi_Hex_Search} = NS_{Multi_Hex_Search} \cdot CS_{Multi_Hex_Search} \\ C_{Small_Hex_Search} = \begin{cases} Let\ it\ go & if\ (NS_{Cross_Search} + NS_{Multi_Hex_Search}) > 1 \\ 0 & if\ (NS_{Cross_Search} + NS_{Multi_Hex_Search}) \le 1 \end{cases} \\ C_{Small_Diamond_Search} = \begin{cases} Let\ it\ go & if\ \ NS_{Cross_Search} > 1 \\ 0 & if\ \ NS_{Cross_Search} \le 1 \end{cases} \end{cases} \quad (8)$$

where $C_{Small_Local_Search}$, $C_{Cross_Search}$, $C_{Multi_Hex_Search}$, $C_{Small_Hex_Search}$ and $C_{Small_Diamond_Search}$ are the computation allocated to the each ME step of the SHS algorithm. $C_{Step_min}$ is the minimum guaranteed computation for the *Small Local Search Step*. In the case of the SHS method, $C_{Step_min}$ is set to be 4. $CS_{Cross_Search}$ and $CS_{Multi_Hex_Search}$ are the number of SPs in each sub-step of the *Cross Search Step* and the *Multi Hexagon Search Step*, respectively. For the SHS method, $CS_{Cross_Search}$ and $CS_{Multi_Hex_Search}$ are equal to 4 and 16, respectively [7]. "*Let it go*" in Eqn. (8) means performing the regular motion search step. $NS_{Cross_Search}$ and $NS_{Multi_Hex_Search}$ are the number of sub-steps in the *Cross Search Step* and the *Multi Hexagon Search Step*, respectively. They are calculated as in Eqn. (9).

$$\begin{cases} NS_{Cross_Search} = \left\lfloor \dfrac{RT_{Cross_Search} \cdot (C_{cur_MB} - C_{Step_min})}{CS_{Cross_Search}} \right\rfloor \\ NS_{Multi_Hex_Search} = \left\lfloor \dfrac{RT_{Multi_Hex_Search} \cdot (C_{cur_MB} - C_{Step_min})}{CS_{Multi_Hex_Search}} \right\rfloor \end{cases} \quad (9)$$

where $C_{cur_MB}$ is the computation budget for the whole MB as in Eqn. (6). $RT_{Cross_Search}$ and $RT_{Multi_Hex_Search}$ are the pre-defined ratios by which the MB's budget $C_{cur_MB}$ is allocated to the *Cross Search Step* and the *Multi Hexagon Search Step*. In the case of SHS method, we set $RT_{Cross_Search}$ to be 0.32 and $RT_{Multi_Hex_Search}$ to be 0.64. This means that 32% of the MB's budget will be allocated to the *Cross Search Step* and 64% of the MB's budget will be allocated to the *Cross Search Step*. We use the floor function ($\lfloor \cdot \rfloor$) in order to make sure that the integer sub-steps of search points are allocated.

From Eqn. (8), we can see that the SLA process will first allocate the minimum guaranteed computation to *the Small Local Search step*. Then most of the available computation budget will be allocated to the *Cross Search Step* (32%) and the *Multi Hexagon Search Step* (64%). If there is still enough computation left after these two steps, the regular *Small Hexagon Search* and *Small Diamond Search* will be performed to refine the final MV. If there is not enough budget for the current MB, some motion search steps such as the *Small Hexagon Search* and *Small Diamond Search* will be skipped. In the extreme case, for example, if the MB's budget only has 6 SPs, then all the steps after the *Small Local Search* will be skipped and the SLA process will end up with only performing a *Small Local Search*. It should be noted that since the SLA is proceeded before the ME process, the computation will be allocated to the *Cross Search* and the *Multi Hexagon Search Steps* no matter whether these steps are skipped in the later ME process (i.e., skipped by $th_2$ in Fig. 1).

# IV. Experimental Results

We implemented our proposed CCME algorithm on the H.264/MPEG-4 AVC reference software JM10.2 version [8]. Motion search was based on Simplified Hexagon Search (SHS) [7] where $th_1$ and $th_2$ in Fig. 1 is set to be *1000* and *5000*, respectively. For each of the sequences, 100 frames were coded, and the picture coding structure was IPPP…. It should be noted that the first P frame was coded by the original SHS method [7] to obtain initial information for each class. In the experiments, only the 16x16 partition was used with one reference frame coding for the P frames. The QP was set to be 28, and the search range was $\pm 32$ pixels.

## A. Experimental results for the CCME Algorithm

In this section, we show experimental results for our proposed CCME algorithm. We fix the target computation (or SP) budget for each frame. The results are shown in Table 3 and Fig. 7.

Table 3 shows *PSNR*, Bit Rate, *the average number of search points actually used per frame* (*Actual SP*) and *the average number of search points per MB* (*Actual SP/MB*) for different sequences. The *Budget* columns in the table represent the target SP budget for performing ME where *100%* in the *Scale* column represents the original SHS [7]. Since we fix the target SP budget for each frame, the values in the *Scale* column are measured in terms of *the number of SPs per frame* (e.g., 40% in the *Scale* column means the target SP budget *for each frame* is 40% of the average-SP-per-frame value of the original SHS [7]).

Similarly, the values in the *Budget SP* column represent the corresponding number of SPs per frame for the budget scale levels indicated by the *Scale* column. Fig. 7 shows the number of SPs used for each frame as well as the target SP budgets for each frame under 60% budget levels for *Football_Cif*. Similar results can be found for other sequences.

Table 3
Experimental results for the Tight Strategy when fixing the target budget for each frame (note: the *Budget SP* and the *Actual SP* columns are measured in terms of the number of SPs per frame)

| Sequence | Budget | | Actual SP | PSNR (dB) | Bit Rate (kbps) | Actual SP/MB |
|---|---|---|---|---|---|---|
| | Scale | Budget SP | | | | |
| Football_Cif | 100% | 22042 | 22042 | 35.96 | 1661.62 | 55 |
| | 60% | 13225 | 10692 | 35.96 | 1678.38 | 27 |
| | 40% | 8816 | 8615 | 35.96 | 1682.57 | 21 |
| Mobile_Cif | 100% | 9871 | 9871 | 33.69 | 2150.60 | 24 |
| | 60% | 5922 | 5785 | 33.69 | 2152.56 | 15 |
| | 40% | 3948 | 3825 | 33.68 | 2165.31 | 10 |

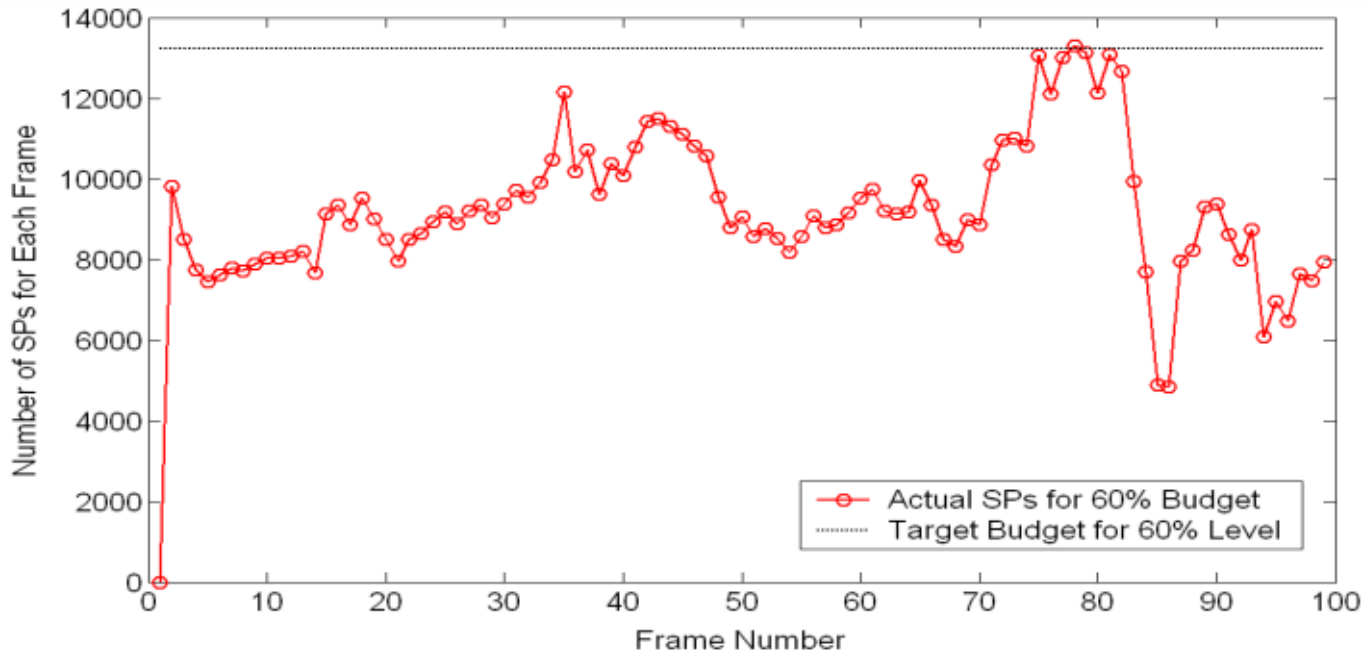

Fig. 7 The number of SPs used for each frame vs. the target frame-level budgets for the tight strategy for Football_Cif.

Comparing the *Actual SP* column with the *Budget SP* column in Table 3, we can see that *the number of SPs actually used* is always smaller than the *target SP budget* for all target budget levels. This demonstrates that our CCME algorithm can efficiently perform computation allocation to meet the requirements of different target computation budgets. From Table 3, we can also see that our CCME algorithm has good performance even when the available budget is low (40% for Football and Mobile). This demonstrates the allocation efficiency of our algorithm. Furthermore, from Fig. 7, we can see that since the CCME algorithm is based on the *tight strategy* which does not allow computation borrowing from other frames, the number of SPs used in each frame is always smaller than the target frame-level budget. Thus, the average SPs per frame for the *tight strategy* is always guaranteed to be smaller than the target budget.

### *B. Comparison with other methods*

In the previous sections, we have shown experimental results for our proposed CCME algorithm. In this section, we will compare our CCME methods with other methods.

Similar to the previous seciton, we fixed the target computation budget for each frame to prevent the effect of frame level allocation. The following three methods are compared. It should be noted that all these three methods use our step-level allocation method for a fair comparison.

**(I)** Perform the proposed CCME algorithm with the *tight strategy* (*Proposed* in Table 4).

**(II)** Do not classify the MBs into classes and allocate computation only based on their *Init_COST* [5,14] (*COST only* in Table 4).

**(III)** First search the (0,0) points of all the MBs in the frame, and then allocate SPs based on (0,0) SAD. This method is the variation of the strategy for many multi-pass methods [2-3] ((*0,0*) *SAD* in Table 4).

Table 4 compares PSNR (in dB), Bit Rate (BR, in kbps), and the average number of search points per MB (SPs). The definition of the *Budget Scale* column of the table is the same as in Table 3. Fig. 8 shows the *BR Increase* vs. *Budget Level* for these methods where the *BR Increase* in defined by the ratio between the current bit-rate and its corresponding 100% Level bit-rate.

From Table 4 and Fig. 8, we can see that our proposed CCME method can allocate SPs more efficiently than the other methods at different computation budget levels. This demonstrates that our proposed method, which combines the class and the COST information of the MB, can provide a more accurate way to allocate SPs.

Table 4
Performance Comparison for CCME algorithms (all sequences are Cif)

| | | Proposed | | | COST Only | | | (0,0) SAD | | |
|---|---|---|---|---|---|---|---|---|---|---|
| | Budget | PSNR | BR | SPs | PSNR | BR | SPs | PSNR | BR | SPs |
| Bus | 100% | 34.31 | 1424 | 35 | 34.31 | 1424 | 35 | 34.31 | 1424 | 35 |
| | 60% | 34.31 | 1459 | 20 | 34.29 | 1484 | 19 | 34.29 | 1482 | 20 |
| | ***40%*** | ***34.29*** | ***1524*** | ***13*** | ***34.25*** | ***1628*** | ***12*** | ***34.27*** | ***1642*** | ***13*** |
| Mobile | 100% | 33.69 | 2151 | 24 | 33.69 | 2151 | 24 | 33.69 | 2151 | 24 |
| | 50% | 33.68 | 2153 | 12 | 33.69 | 2187 | 12 | 33.69 | 2196 | 11 |
| | ***30%*** | ***33.68*** | ***2167*** | ***7*** | ***33.66*** | ***2276*** | ***7*** | ***33.66*** | ***2283*** | ***7*** |
| Stefan | 100% | 35.12 | 1354 | 22 | 35.12 | 1354 | 22 | 35.12 | 1354 | 22 |
| | 50% | 35.11 | 1369 | 11 | 35.09 | 1404 | 10 | 35.09 | 1394 | 11 |
| | ***35%*** | ***35.10*** | ***1376*** | ***7*** | ***34.98*** | ***1703*** | ***7*** | ***35.05*** | ***1642*** | ***7*** |
| Dancer | 100% | 39.09 | 658 | 16 | 39.09 | 658 | 16 | 39.09 | 658 | 16 |
| | 60% | 39.10 | 701 | 9 | 39.12 | 746 | 9 | 39.11 | 732 | 8 |
| | ***50%*** | ***39.10*** | ***717*** | ***8*** | ***39.11*** | ***768*** | ***7*** | ***39.12*** | ***756*** | ***7*** |
| Foreman | 100% | 36.21 | 515 | 16 | 36.21 | 515 | 16 | 36.21 | 515 | 16 |
| | 70% | 36.21 | 520 | 11 | 36.21 | 519 | 10 | 36.22 | 520 | 10 |
| | ***50%*** | ***36.22*** | ***522*** | ***8*** | ***36.21*** | ***522*** | ***7*** | ***36.22*** | ***523*** | ***8*** |
| Football | 100% | 35.96 | 1662 | 55 | 35.96 | 1662 | 55 | 35.96 | 1662 | 55 |
| | 60% | 35.96 | 1678 | 27 | 35.96 | 1681 | 29 | 35.97 | 1689 | 28 |
| | ***40%*** | ***35.96*** | ***1682*** | ***21*** | ***35.95*** | ***1719*** | ***21*** | ***35.96*** | ***1711*** | ***21*** |

For a further analysis of the result, we can compare the bit-rate performance of the Mobile sequence (i.e., Fig. 8 (b)) with its MB classification result (i.e., Fig. 3 (b)-(d)). When the budget level is low, our proposed algorithm can efficiently extract and allocate more SPs to the more important Class 2 MBs (Fig. 3 (c)) while reducing the unnecessary SPs from Class 3 (Fig. 3 (d)). This keeps the performance of our method as high as possible. Furthermore, since the number of extracted Class 2 MBs is low (Fig. 3 (c)), our proposed algorithm can still keep high performance at very low budget levels (e.g., 5% budget level in Fig. 8 (b)). Compared to our method, the performances of the other methods will significantly decrease when the budget level becomes low.

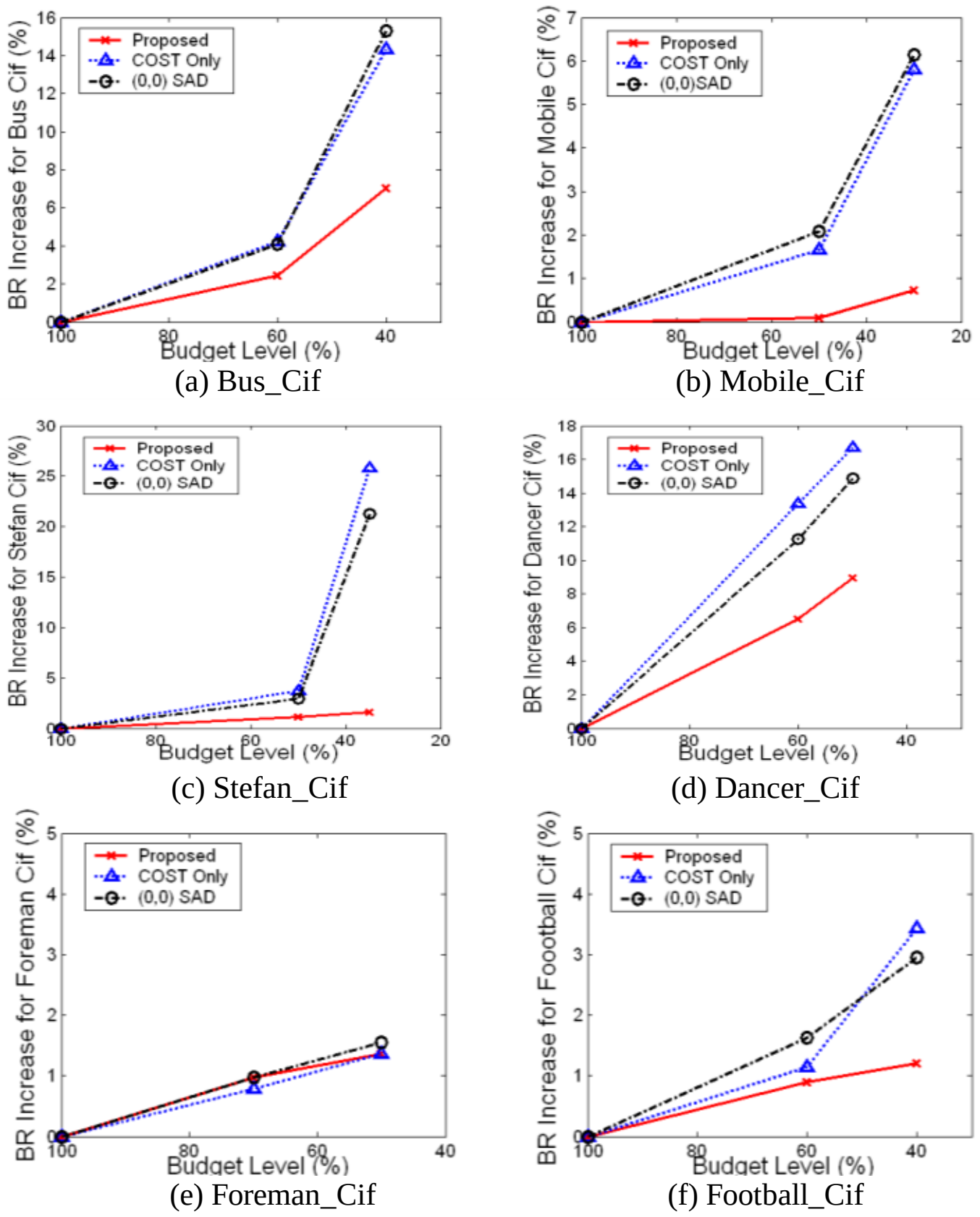

(a) Bus_Cif (b) Mobile_Cif

(c) Stefan_Cif (d) Dancer_Cif

(e) Foreman_Cif (f) Football_Cif

Fig. 8 Performance comparison for different CCME algorithms.

However, the results in Table 4 and Fig. 8 also show that for some sequences (e.g., Foreman and Football), the advantage of our CCME algorithm are not so obvious from the other methods. This is because:

(1) For some sequences such as *Football*, the portion of Class 2 MBs is large. In this case, the advantages of our CCME method from MB classification become less obvious. In extreme cases, if all MBs are classified into Class 2, our proposed CCME algorithm will be the same as the *COST only* algorithm).

(2) For some sequences such as *Foreman*, the performance will not decrease much even when very few points are searched for each MB (e.g., our experiments show that the performance for *Foreman_Cif* will not decrease much even if we only search 6 points for each MB). In this case, different computation allocation strategies will not make much difference.

Table 5 shows the results for sequences with different resolutions (Mobile_Qcif and Mobile_SD) or using different QPs (Bus with QP =23 or 33). Table 5 shows the efficiency of our algorithm under different resolutions and different QPs. Furthermore, we can also see from Table 5 that the performance of our algorithm is very close to the other methods for Mobile_Qcif. The reason is similar to the case of Foreman_Cif (i.e., a local search for each MB can still get good performance and thus different computation allocation strategies will not make much difference).

Table 5
Experimental results for sequences with different resolutions or different QPs.

| | Budget | Proposed | | | COST Only | | | (0,0) SAD | | |
|---|---|---|---|---|---|---|---|---|---|---|
| | | PSNR | BR | SPs | PSNR | BR | SPs | PSNR | BR | SPs |
| Bus Cif QP=23 | 100% | 38.28 | 2639 | 33 | 38.28 | 2639 | 33 | 38.28 | 2639 | 33 |
| | ***50%*** | ***38.26*** | ***2762*** | ***14*** | ***38.23*** | ***2912*** | ***13*** | ***38.24*** | ***2896*** | ***14*** |
| Bus Cif QP=33 | 100% | 30.47 | 722 | 40 | 30.47 | 722 | 40 | 30.47 | 722 | 40 |
| | ***50%*** | ***30.46*** | ***789*** | ***16*** | ***30.41*** | ***902*** | ***15*** | ***30.41*** | ***879*** | ***15*** |
| Mobile Qcif QP=28 | 100% | 32.90 | 545 | 16 | 32.90 | 545 | 16 | 32.90 | 545 | 16 |
| | ***50%*** | ***32.90*** | ***545*** | ***7*** | ***32.90*** | ***546*** | ***7*** | ***32.90*** | ***545*** | ***7*** |
| M obile SD QP=28 | 100% | 34.07 | 7766 | 24 | 34.07 | 7766 | 24 | 34.07 | 7766 | 24 |
| | ***30%*** | ***34.07*** | ***7776*** | ***7*** | ***34.06*** | ***8076*** | ***7*** | ***34.05*** | ***8124*** | ***7*** |

## V. Discussion and Algorithm Extension

The advantages of our proposed CCME algorithm can be summarized as follows:

(1) The proposed algorithm uses a *more suitable way to measure MB importance* by differentiating MBs into different classes. When the available budget is small, the proposed method can save unnecessary SPs from Class 3 MBs so that more SPs can be allocated to the more important Class 2 MBs, which keeps the performance as high as possible. When the available target budget is large, the method will have more spare SPs for Class 3 MBs, which can overcome the possible performance decrease from MB mis-classification and further improve the coding performance.

(2) The proposed algorithm can reduce the impact of not having a global view of the whole frame for one-pass methods by (i) setting the basic and the additional layers, (ii) using previous frame information as the global view estimation, (iii) guaranteeing Class 2 MBs a higher minimum SPs, and (iv) using three independent class budgets so that an unsuitable allocation in one class will not affect other classes.

Furthermore, we also believe the framework of our CCME algorithm is general and can easily be extended. Some possible extensions of our algorithm can be described as follows:

(1) As mentioned, other FLA or SLA methods [1-5,14] can easily be implemented into our CCME algorithm. For example, in some time-varying motion sequences, an FLA algorithm may be very useful to allocate more computation to those high-motion frames and further improve the performance.

(2) In this paper, we only perform experiments on the 16x16 partition size and the IPPP… picture type. Our algorithm can easily be extended to ME with multiple partition sizes as well as multiple reference frames such as in H.264|AVC [12] as well as other picture types.

(3) In this paper, we define three MB classes and perform CCME based on these three classes. Our method can also be extended by defining more MB classes and developing different CLA and MLA steps for different classes.

## VI. Conclusion

In this paper, we propose a more accurate MB Importance Measure method by introducing the definition of class. A new one-pass CCME is then proposed based on the new measure method. The four computation allocation steps of FLA, CLA, MLA, and SLA in the proposed CCME algorithm are introduced in the paper. Experimental results demonstrate that the proposed method can allocate computation more accurately and efficiently than previous methods to achieve better coding performance.